%% file: main.tex
\documentclass[sigconf]{acmart}

\usepackage{multirow}
\usepackage{enumitem}
\usepackage{subfigure}

\usepackage[utf8]{inputenc} 
\usepackage[T1]{fontenc}    
\usepackage{hyperref}       
\usepackage{url}            
\usepackage{amsfonts}       
\usepackage{microtype}      
\usepackage{xcolor}         
\usepackage{multirow}       
\usepackage[ruled,linesnumbered]{algorithm2e}
\usepackage{graphicx}
\usepackage{amsmath}
\usepackage{amsthm}

\newtheorem{definition}{Definition}

\usepackage{appendix}
\usepackage{bm}
\usepackage{enumitem}
\usepackage{mathtools} 
\usepackage{fancyhdr}

\AtBeginDocument{%
  \providecommand\BibTeX{{%
    \normalfont B\kern-0.5em{\scshape i\kern-0.25em b}\kern-0.8em\TeX}}}

\copyrightyear{2023}
\acmYear{2023}
\setcopyright{acmlicensed}\acmConference[WWW '23]{Proceedings of the ACM Web Conference 2023}{April 30--May 4, 2023}{Austin, TX, USA}
\acmBooktitle{Proceedings of the ACM Web Conference 2023 (WWW '23), April 30--May 4, 2023, Austin, TX, USA}
\acmPrice{15.00}
\acmDOI{10.1145/3543507.3583237}
\acmISBN{978-1-4503-9416-1/23/04}

\begin{document}

\title{Dynamically Expandable Graph Convolution for Streaming Recommendation}

\author{Bowei He}
\authornote{Work done as an intern in Huawei Noah's Ark Lab, Hong Kong.}
\affiliation{%
  \institution{City University of Hong Kong}
  \country{Hong Kong SAR}
}
\email{boweihe2-c@my.cityu.edu.hk}

\author{Xu He}
\affiliation{%
  \institution{Huawei Noah's Ark Lab}
  \city{Shenzhen}
  \country{China}
  }
\email{hexu27@huawei.com}

\author{Yingxue Zhang}
\affiliation{%
  \institution{Huawei Noah's Ark Lab Montreal}
  \city{Montreal}
  \country{Canada}
}
\email{yingxue.zhang@huawei.com}

\author{Ruiming Tang}
\affiliation{%
 \institution{Huawei Noah's Ark Lab}
 \city{Shenzhen}
 \country{China}
 }
 \email{tangruiming@huawei.com}

\author{Chen Ma}
\authornote{Corresponding author}
\affiliation{%
  \institution{City University of Hong Kong}
  \country{Hong Kong SAR}}
\email{chenma@cityu.edu.hk}

\renewcommand{\shortauthors}{Bowei He et al.}

\begin{abstract}
  Personalized recommender systems have been widely studied and deployed to reduce information overload and satisfy users' diverse needs. However, conventional recommendation models solely conduct a one-time training-test fashion and can hardly adapt to evolving demands, considering user preference shifts and ever-increasing users and items in the real world. To tackle such challenges, the streaming recommendation is proposed and has attracted great attention recently. Among these, continual graph learning is widely regarded as a promising approach for the streaming recommendation by academia and industry. However, existing methods either rely on the historical data replay which is often not practical under increasingly strict data regulations, or can seldom solve the \textit{over-stability} issue. To overcome these difficulties, we propose a novel \textbf{D}ynamically \textbf{E}xpandable \textbf{G}raph \textbf{C}onvolution (DEGC) algorithm from a \textit{model isolation} perspective for the streaming recommendation which is orthogonal to previous methods. Based on the motivation of disentangling outdated short-term preferences from useful long-term preferences, we design a sequence of operations including graph convolution pruning, refining, and expanding to only preserve beneficial long-term preference-related parameters and extract fresh short-term preferences. Moreover, we model the temporal user preference, which is utilized as user embedding initialization, for better capturing the individual-level preference shifts. Extensive experiments on the three most representative GCN-based recommendation models and four industrial datasets demonstrate the effectiveness and robustness of our method.
\end{abstract}



\maketitle
\input{introduction.tex}

\input{related_work.tex}

\input{preliminaries.tex}

\input{methodology.tex}

\input{experiments.tex}

\input{conclusion.tex}

\begin{acks}
This work was supported by the Start-up Grant (No. 9610564) and the Strategic Research Grant (No. 7005847) of City University of Hong Kong.
\end{acks}

\bibliographystyle{ACM-Reference-Format.bst}
\bibliography{reference.bib}
\appendix
\input{appendix.tex}


\end{document}

%% file: introduction.tex
\section{Introduction}
\label{sec:introduction}
\begin{figure}[ht]
    \centering
    \includegraphics[width=0.5\textwidth]{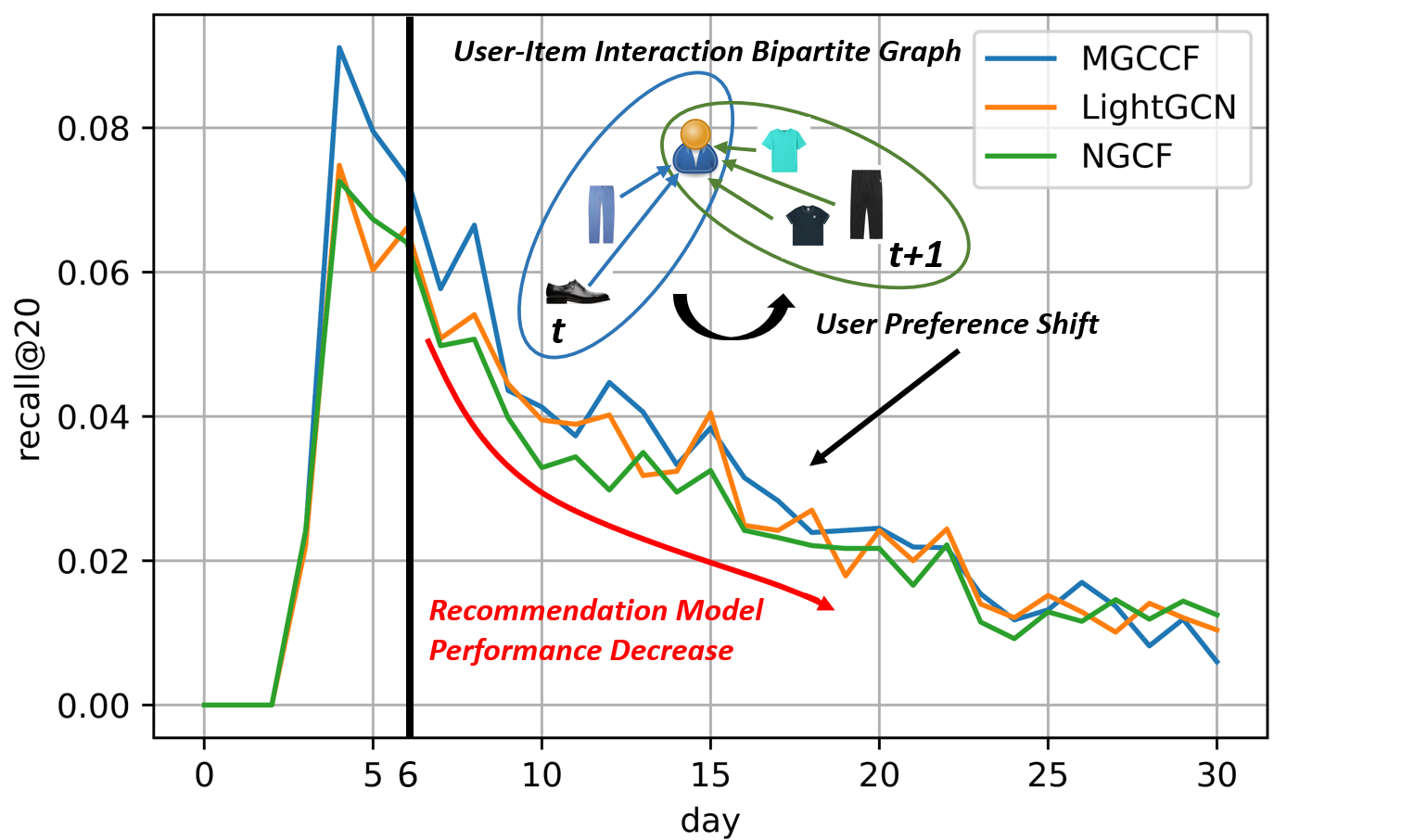}
    \caption{LightGCN, NCGF, and MGCCF only trained in the first week and then tested in the following three weeks on Taobao2014 dataset.
    }
    \Description{LightGCN, NCGF, and MGCCF only trained in the first week and then tested in the following three weeks on Taobao2014 dataset.}
    \label{fig:decrease}
    \vspace{-0.3cm}
\end{figure}

The recommender system (RS), aiming to provide the personalized contents to different users precisely~\citep{ma2019hierarchical, ma2020probabilistic, chen2021attentive, chen2021hyper, chen2022learning}, has been deployed in many online Internet applications. However, traditional recommender systems trained on the offline static datasets face three challenges in the real online platforms: user preference shift, ever-increasing users and items, and intermittent user activities. In fact, these can cause the unacceptable performance degradation as shown in Figure~\ref{fig:decrease}, and lead to the necessity of dynamical model updating. Thus, how to update the model dynamically to tackle the above challenges has attracted great attention in real-world RS~\citep{cai2022reloop, sima2022ekko}. This ensures the indispensable need of \textit{streaming recommendation}, referring to updating and applying recommendation models dynamically over the data stream. 

Due to the strong capability of modeling user-item, user-user, and item-item relationships, graph convolution neural network (GCN) has been widely-used as recommendation models. Some recent works~\citep{xu2020graphsail, wang2021graph, ahrabian2021structure, wang2020streaming, wang2022streaming} develop continual learning methods for GCN-based recommendation methods to achieve the streaming recommendation, also known as continual graph learning for streaming recommendation.

To enable continual GCN-based recommendation, most works focus on two realizations: \textit{experience replay}~\citep{ahrabian2021structure, zhou2021overcoming, wang2022streaming} and \textit{knowledge distillation/weight regularization}~\citep{wang2021graph, liu2020overcoming, xu2020graphsail, wang2020streaming}. Although these methods have achieved acceptable results, there are still drawbacks which hinder them from being applied in the real-world systems. First, \textit{experience replay} needs complete and accurate historical data when training the model on the newly coming data. Nevertheless, in the real-world system, the missing data~\citep{chi2020missing, wang2019doubly} and malicious data attack~\citep{zhang2021data} issues are pretty common. What's worse, more and more strict data governance policies often make the user historical behavior inaccessible. Training with historical data replay also brings the huge memory cost and increasing time consumption. Second, knowledge distillation/model regularization can hardly capture the varying short-term preferences, especially for those users whose preferences change quickly and dramatically. Such issue is also known as \textit{over-stability}~\citep{ostapenko2021continual} in continual learning. Third, previous continual graph learning methods for streaming recommendation do not model the user-level temporal preference changes explicitly and only relies on the graph convolution itself to capture the shift, which is greatly limited by the aforementioned two challenges: ever-increasing users and intermittent user activities. This is also far from the fine-grained user modeling for personalized recommendation. Besides, few existing \textit{model isolation}-based methods~\citep{cai2022multimodal, yuan2021one} (another main approach for continual learning) are designed specially for the typical continual learning paradigm with task boundaries and can hardly handle the growing dataset (data stream) on a single task. Such \textit{model isolation} methods need to search the whole GCN architecture in each task which increases the method complexity greatly. In streaming recommendation setting, \textit{model isolation}-based continual graph learning is less investigated though some apparent advantages like no longer needing to replay historical data and the potential to overcome the \textit{over-stability} issue.


To tackle the above challenges, we propose a model-isolation continual graph learning method, namely \textbf{D}ynamically \textbf{E}xpandable \textbf{G}raph \textbf{C}onvolution (DEGC), to better model the user preference shift without the need of historical data replay. First, we design a graph convolution network-based sparsification training method to disentangle short-time preference-related parameters from long-time preference-related parameters. Then we remove outdated short-term preference-related filters and preserve long-term preference-related filters which are further refined with newly-collected data. Next, the graph convolution network is expanded by additional filters to extract the current short-term preference. The added filters will also be partly pruned to eliminate the redundant ones and prevent the network width explosion catastrophe. Moreover, inspired by the Kalman filter~\citep{welch1995introduction}, a temporal attention model is utilized to explicitly encode the temporal user preference, which works as the user embedding initialization for training on new data.

In summary, the main contributions of this paper are:
 \begin{itemize}[leftmargin=*]
   \item We propose a \textit{model isolation}-based continual graph learning method, DEGC, for streaming recommendation. We design a sequence of graph convolution operations including pruning, refining, and expanding to overcome the \textit{over-stability} challenge.  
   
   \item To address the challenges of ever-increasing users and intermittent user activities, we model the temporal user preference as the user embedding initialization to help learn users' preferences on newly coming data. 
   
   \item Experiments on three representative GCN-based recommendation models and four real-world datasets demonstrate the effectiveness and robustness of our method.
 \end{itemize}

%% file: related_work.tex
\section{Related Work}
\subsection{Streaming Recommendation}
Due to the real-world dynamics like user preference continuous shift and ever-increasing users and items, conventional recommender systems trained on the static fixed datasets usually suffer from: predicting previous interactions and preferences, disregarding trends and shifting preferences, and ignoring real industrial constraints like few time and limited resources. To tackle these challenges, streaming recommendation is proposed in which data and recommendation model are both updated dynamically along the timeline~\citep{das2007google, song2008real, chen2013terec, song2017multi, devooght2015dynamic, wang2018streaming, chang2017streaming}. Early works recommend items to users based on the popularity, recency, and trend analysis~\citep{chandramouli2011streamrec,lommatzsch2015real, subbian2016recommendations} but pay few attention to the collaborative signal distilling. To extracting such information, some other works~\citep{rendle2008online, diaz2012real, devooght2015dynamic, chang2017streaming} introduce the classical recommendation algorithms like collaborative filtering and machine factorization into the streaming setting. In addition, there are also some recent works from the perspectives of online clustering of bandits and collaborative filtering bandits~\citep{ban2021local, shuai2019online, gentile2017context, gentile2014online, li2016collaborative} to perform streaming recommendation. Thanks to the great success of graph neural network on complex relationship modeling, how to apply GCN-based recommendation models to the streaming recommendation is attracting more and more attention recently~\citep{xu2020graphsail, wang2021graph, ahrabian2021structure, wang2020streaming, wang2022streaming}.  Besides, streaming recommendation algorithms have been successfully deployed to industrial online service platforms like Google, Huawei, and Tencent~\citep{das2007google, cai2022reloop, sima2022ekko}. However, for a long time, there lacks a standardized definition to streaming recommendation, especially in the deep model-based recommendation setting. In this paper, we draw intuitions from previous research and most recent progress, and then summarize a definition of the streaming recommendation.

 
\subsection{Continual Learning}
Continual learning was originally paid great attention in computer vision and nature language process areas in which different tasks come in sequence. Various methods have been proposed to prevent catastrophic forgetting and effectively transfer knowledge. The mainstream continual learning algorithms can be classified into three categories: \textit{experience replay}~\citep{chaudhry2019continual, isele2018selective, prabhu2020gdumb, rebuffi2017icarl, mi2020ader}, \textit{knowledge distillation/model regularization}~\citep{hinton2015distilling, rannen2017encoder, dhar2019learning, hou2019learning}, and \textit{model isolation}~\citep{xu2018reinforced, DEN,rusu2016progressive, ostapenko2021continual, wu2020firefly, qin2021bns, golkar2019continual}. Continual learning is often regarded as a trade-off between knowledge retention (stability) and knowledge expansion (plasticity)~\citep{ostapenko2021continual}, and \textit{model isolation}-based methods provide an more explicit control over such trade-off. Considering that graph-based models have been widely studied to model the complex data relationships, continual graph learning~\citep{ma2020streaming, wang2020streaming, wang2022streaming, EvolveGCN, liu2020overcoming, zhou2021overcoming, perini2022learning, cai2022multimodal} has also attracted more and more attentions recently. When it comes to the continual graph in recommendation setting~\citep{xu2020graphsail, wang2021graph, ahrabian2021structure, wang2020streaming}, we focus more on the data coming continuously in chronological order rather than the data with task boundaries. Different from the conventional continual learning, the continual graph learning for streaming recommendation which is studied in this work pays more attention to the effective knowledge transfer across the time segments rather than preventing catastrophic forgetting. This is because performance degradation on historical data makes no sense to a real recommender system.

%% file: preliminaries.tex
\section{Preliminaries}
\label{section:preliminaries}
In this section, we first formalize the continual graph learning for streaming recommendation. Then we briefly introduce three classical graph convolution based recommendation models used in this paper.

\subsection{Definitions and Formulations}
\begin{definition}{Streaming Recommendation.}
Massive user-item interaction data $\widetilde{\textbf{D}}$ streams into industrial recommender system continuously. For convenience~\citep{caccia2021anytime, wang2020streaming, wang2022streaming}, the continuous data stream is split into consecutive data  segments $D_1,...,D_t, ..., D_T$ with the same time span. At each time segment $t$, the model needs to optimize the recommendation performance on $D_t$ with the knowledge inherited from $D_1, D_2,..., D_{t-1}$. The recommendation performance is evaluated along the whole timeline.
\end{definition}

\begin{definition}{Streaming Graph.}
A streaming graph is represented as a sequence of graphs $\mathcal{G}= (G_1, G_2, ..., G_t, ..., G_T)$, where $G_t = G_{t-1} + \Delta G_t$. $G_t = (\mathbf{A}_t, \mathbf{X}_t)$ is an attributed graph at time $t$, where $\mathbf{A}_t$ and $\mathbf{X}_t$ are the adjacency matrix and node features of $G_t$, respectively. $\Delta G_t = (\Delta \mathbf{A}_t, \Delta \mathbf{X}_t)$ is the changes of graph structures and node attributes at $t$. The changes contain newly added nodes and newly built connections between different nodes.
\end{definition}

\begin{definition}{Continual Graph Learning for Streaming Graph.}
\label{def:cgl}
Given a streaming graph $\mathcal{G}= (G_1, G_2, ..., G_t, ..., G_T)$, the goal of \textbf{continual graph learning (CGL)} is to learn $\Delta G_t (D_t)$ sequentially while transferring historical knowledge to new graph segments effectively. Mathematically, the goal of \textbf{CGL} for streaming graph is to find the optimal GNN structure $\mathbf{S}_t$ and parameters $\mathbf{W}_t$ at each segment $t$ such that:
\begin{equation}
    (\mathbf{S}^{*}_t, \mathbf{W}^{*}_t) = \mathop{\arg\min}\limits_{(\mathbf{S}_t, \mathbf{W}_t)}\mathcal{L}_t(\mathbf{S}_t, \mathbf{W}_t, \Delta G_t),
\end{equation}
where $(\mathbf{S}_t, \mathbf{W}_t) \in (\mathcal{S}, \mathcal{W}) $. $\mathcal{L}_t(\mathbf{S}_t, \mathbf{W}_t, \Delta G_t)$ is the loss function of current task defined on $\Delta G_t$. The $\mathcal{S}$ and $\mathcal{W}$ are corresponding search spaces, respectively.
\end{definition}

Since the user-item interaction data is actually a bipartite graph, the continual learning task for streaming recommendation is essentially the continual graph learning for streaming graph. For each segment $t$, the GNN structure $\mathbf{S}_t$ and parameters $\mathbf{W}_t$ need to be adjusted and refined simultaneously to achieve a satisfying recommendation performance. We use the Bayesian Personalized Ranking (BPR)~\citep{bprloss} loss as the loss function in this work, because it is effective and has broad applicability in top-K recommendation tasks. The major notations are summarized in Appendix \ref{appendix:notations}.


\subsection{GCN-based Recommender Models}
Many graph convolution-based recommender models~\citep{wang2019neural, he2020lightgcn, sun2019multi,ying2018graph} have been developed recently to capture the collaborative signal, which is not encoded by the early matrix factorization and other deep learning based models. A general graph convolution process for such models can be summarized below: 
On the user-item bipartite graph, the layer-$k$ embedding of user $u$ is obtained via the following processing: 
\begin{equation}
\begin{aligned}
\mathbf{h}^{u,k} &= \sigma(\mathbf{W}^{u,k} \cdot [\mathbf{h}^{u,k-1}; \mathbf{h}^{\mathcal{N}(u),k-1}]), \mathbf{h}^{u,0} = \mathbf{e}^u,\\
\mathbf{h}^{\mathcal{N}(u),k-1} &= AGGREGATOR^u({\mathbf{h}^{i,k-1}, i \in \mathcal{N}(u)}), \\
\end{aligned}
\end{equation}
where $\mathbf{e}^u$ is the initial user embeddings, $\sigma(\cdot)$ is the activation function, $\mathbf{h}^{\mathcal{N}(u),k-1}$ is the learned neighborhood embedding, and $\mathbf{W}^{u,k}$ is the layer-$k$ user transformation matrix shared among all users. The $AGGREGATOR^u$ is the designed aggregation function in order to aggregate neighbor information for user nodes. Similarly, the layer-$k$ embedding of item $i$ is obtained via the following processing:
\begin{equation}
\begin{aligned}
\mathbf{h}^{i,k} &= \sigma(\mathbf{W}^{i,k} \cdot [\mathbf{h}^{i,k-1}; \mathbf{h}^{\mathcal{N}(i),k-1}]), \mathbf{h}^{i,0} = \mathbf{e}^i,\\
\mathbf{h}^{\mathcal{N}(i),k-1} &= AGGREGATOR^i({\mathbf{h}^{u,k-1}, u \in \mathcal{N}(i)}). \\
\end{aligned}
\end{equation}
In our setting, considering the timestamps of the above matrices, convolution parameters $\mathbf{W}_t = \left\{\mathbf{W}^{u, 1}_t, \mathbf{W}^{i, 1}_t,..., \mathbf{W}^{u, k}_t, \mathbf{W}^{i, k}_t,...,\mathbf{W}^{u, K}_t, \mathbf{W}^{i, K}_t\right\}$, where $\mathbf{W}^{u, k}_t$ is the $\mathbf{W}^{u,k}$ on the time segment $t$. Each column in such matrices is a graph convolution filter. NGCF~\citep{wang2019neural}, LightGCN~\citep{he2020lightgcn}, and MGCCF~\citep{sun2019multi} are three representative GCN-based recommendation models which will be utilized in our work. Some of their details are provided below.

\textbf{NGCF~\citep{wang2019neural}}. One of the most widely used GCN-based recommendation model. NGCF exploits the user-item bipartite graph structure by propagating embeddings on it with graph convolution.  

\textbf{LightGCN~\citep{he2020lightgcn}}. An improved version of NGCF which still convolutes on user-item graph. Compared with NGCF, LightGCN no more needs neighbor node feature transformation and nonlinear activation. Based on the original model setting, we add a dense matrix to each layer to align the dimensions of aggregated embeddings for better adapting to our approach.

\textbf{MGCCF~\citep{sun2019multi}}. A graph convolution-based recommender framework which explicitly incorporates multiple graphs in the embedding learning process. Compared with the above two models, MGCCF adds a Multi-Graph Encoding(MGE) module to capture the inter-user and inter-item proximity information from user-user graph and item-item graph via homogeneous graph convolution, respectively.


%% file: methodology.tex
\section{Methodology}
\begin{figure*}[ht]
    \centering
    \includegraphics[width=0.88\textwidth]{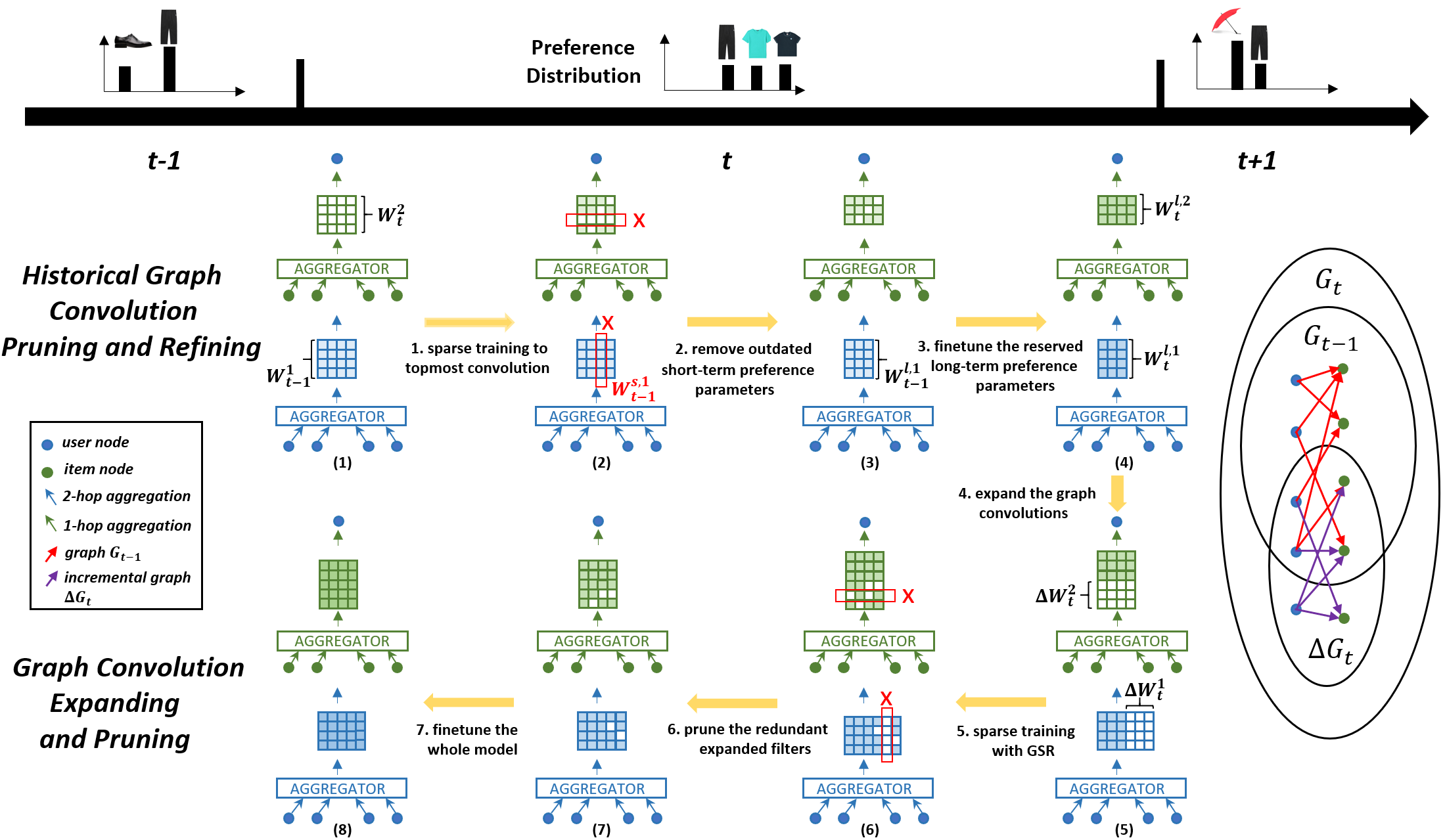}
    \caption{Overview to Dynamically Expandable Graph Convolution. Take 2-layer graph convolution as an example. Operations 1,2,3 correspond to the methods introduced in Sec.~\ref{sec:his}. And operations 4,5,6,7 illustrate the methods mentioned in Sec.~\ref{sec:gra}.
    }
    \Description{Overview to Dynamically Expandable Graph Convolution. Take 2-layer graph convolution as an example. Operations 1,2,3 correspond to the methods introduced in Sec.~\ref{sec:his}. And operations 4,5,6,7 illustrate the methods mentioned in Sec.~\ref{sec:gra}.}
    \label{fig:overview}
    \vspace{-0.3cm}
\end{figure*}

In this section, we introduce our \textbf{DEGC} method towards continual graph learning for streaming recommendation. We first model the temporal user preference to capture the preference shifts between segments, which will be utilized for the user embedding initialization of training. Then, we successively use two operations, \textit{historical graph convolution pruning and refining} as well as \textit{graph convolution expanding and pruning} (shown in Figure \ref{fig:overview}), in such a \textit{model isolation} way to obtain the best structure and optimal parameters of the graph convolution.
\subsection{Temporal User Preference Modeling as Initialization}
\begin{figure}[ht]
    \centering
    \includegraphics[width=0.48\textwidth]{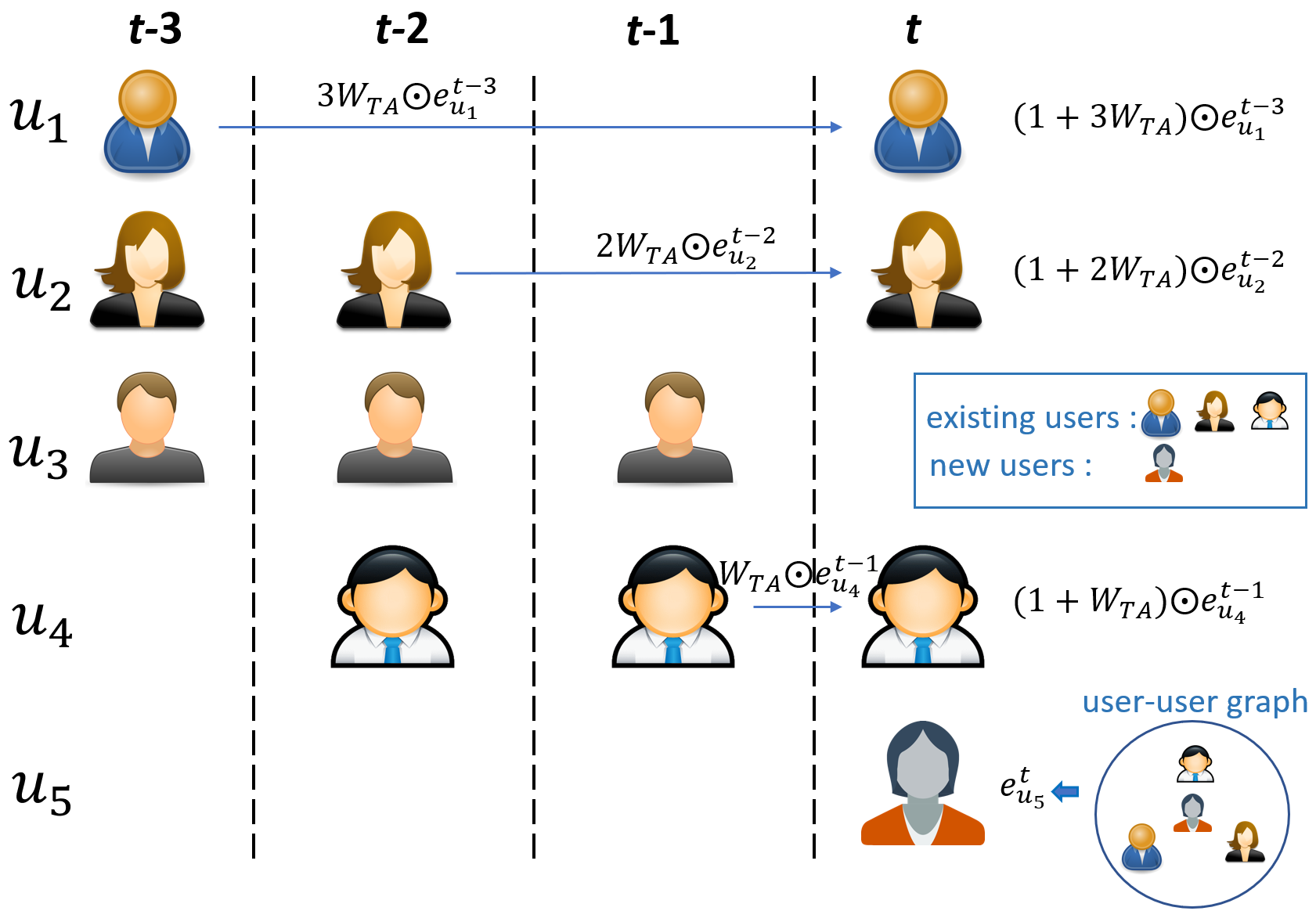}
    \caption{Temporal user preference modeling and new user initialization.
    }
    \Description{Temporal user preference modeling and new user initialization.}
    \label{fig:tam}
    \vspace{-0.3cm}
\end{figure}

Conventional continual graph learning algorithms directly inherit the user embeddings trained in the previous segments as the initialization without considering the user preference shift between the time segments. This type of methods totally rely on the graph convolution network to capture the user preference shift which is far from the fine-grained user modeling. To explicitly model the user-level preference shift and provide a better embedding initialization for graph convolution model training, we propose a temporal attention (TA) module to model the user temporal preference shift. This is motivated by the Kalman filter~\citep{welch1995introduction, kumar2019predicting, beutel2018latent}, which is an effective way to model the temporal state change.  In recommender system, the user embedding $\mathbf{e}^u$ is utilized to represent the user $u$'s preference. At segment $t$, the preference shift of user $u$ can be estimated with the Hadamard product of a time-scaled attention vector and the previous user embedding:
\begin{equation}
\begin{aligned}
\Delta \mathbf{e}^{u}_t &= (\mathbf{W}_{TA} \Delta t) \odot \mathbf{e}^{u}_{t^{-}}, \Delta t = t - t^{-},\\ 
\end{aligned}
\label{equ:5}
\end{equation}
where $t^{-}$ is the last time segment that user $u$ appears and $\mathbf{W}_{TA}$ is a learnable linear matrix. The larger the time interval $\Delta t$, the larger the user preference shift. Note that the users who are intermittently active on the platform (e.g., $u_1$ and $u_2$ in Figure ~\ref{fig:tam}) can also be modeled. Summing the user preference shift vector $\Delta \mathbf{e}^{u}_t$ and the user's previous preference vector $\mathbf{e}^u_{t^{-}}$ gives the user's current preference $\mathbf{e}^{u}_t$:
\begin{equation}
\begin{aligned}
\mathbf{e}^{u}_t &= \mathbf{e}^u_{t^{-}} + \Delta \mathbf{e}^{u}_t.\\ 
\end{aligned}
\label{equ:6}
\end{equation}

As illustrated in Figure~\ref{fig:tam}, the preferences of existing users (have appeared in the historical segments) at segment $t$ can be estimated with the temporal attention module directly. As for new users (appears for the first time), their preference vectors are initialized with a one-hop neighbor aggregation from the user-user graph at segment $t$:
\begin{equation}
\begin{aligned}
 \mathbf{e}^{u}_t &= \frac{1}{|\mathcal{N}^u_t|}\underset{j \in \mathcal{N}^u_t}{\sum} \mathbf{e}^j_t,\\ 
\end{aligned}
\label{equ:7}
\end{equation}
where $j$ are existing users in user $u$'s one-hop neighbor on the user-user graph at segment $t$. The embeddings of existing users and new users will be utilized as the initial embeddings for graph convolution-based recommendation model training.

\subsection{Historical Graph Convolution Pruning and Refining}
\label{sec:his}
In the recommender system, a user's preference is often regarded as the combination of her long-term preference (LTP) and short-term preference (STP)~\citep{rendle2010factorizing, ma2019hierarchical, ma2020probabilistic}. The long-term preference will not change drastically along the time and can take effect at most segments. This kind of preference is often determined by users' gender, occupation, education, and so on. 
By contrast, the short-term preference varies quickly and can only take effect in a certain time segment. STP is often influenced by the recommendation context information, like user emotion. And the recommendation model parameters store both users' long-term preferences
and short-term preferences after learning on the interaction data. Previous continual graph learning methods for streaming recommendation inherit the parameters learned on the last segment indiscriminately and fine-tune them with the new data. However, these methods will preserve the outdated short-term preferences which only work for the last segment and hinder the model learning on the new segment, which corresponds to the \textit{over-stability} issue mentioned above. 

Based on the motivation of decoupling users' useful long-term preferences and outdated short-term preferences, we first design a spasification training method to disentangle LTP-related graph convolution parameters $\mathbf{W}^l_t$ and STP-related graph convolution parameters $\mathbf{W}^s_t$ at segment $t$. As shown in subfigure (1) of Figure~\ref{fig:overview}, we randomly initialize the topmost graph convolution layer $\mathbf{W}^K_t$ and fix the rest graph convolution layers with parameters $\mathbf{W}_{t-1}$ learned on $\Delta G_{t-1}$ and then only train the topmost graph convolution layer $\mathbf{W}^K_t$ with the new incremental interaction graph $\Delta G_{t}$:
\begin{equation}
    \underset{\mathbf{W}^K_t}{min} \quad \mathcal{L}_t(\mathbf{W}^K_t;\mathbf{W}^{1:K-1}_{t-1} , \mathbf{e}, \Delta G_t) + \lambda_1 \lVert \mathbf{W}^K_t \rVert_1.
    \label{equ:8}
\end{equation}
The $\mathbf{e}$ denotes the embeddings of graph nodes. The added $L1$ regulation term is to encourage the sparse connection between layer $K$ and layer $K-1$ illustrated in subfigure (2) of Figure~\ref{fig:overview}. Once obtaining sparse $\mathbf{W}^K_t$, we can identify the filters of $\mathbf{W}^{K-1}_{t-1}$ that have no connection with layer $K$. Starting from this, we can find all the parameters in layer $1: K-1$ that have no connection with convolution layer $K$ via breadth-first search. Because this sparsification effect is obtained by training on $\Delta G_{t}$, such parameters are actually the users' outdated short-term preference-related parameters $\mathbf{W}^s_{t-1}$ and cannot reflect users' current preferences at $t$. And the convolution parameters $\mathbf{W}^l_{t-1}$ that have connection with $\mathbf{W}^K_t$ represent another part of users' previous preferences that still takes effect at $t$, which is users' long-term preferences. Considering that the users' long-term preferences are also evolving along the time, we first remove the $\mathbf{W}^s_{t-1}$ to avoid the negative knowledge transfer and then fine-tune the remaining LTP-related graph convolution parameters $\mathbf{W}^l_{t}$ initialized with $\mathbf{W}^l_{t-1}$:
\begin{equation}
    \underset{\mathbf{W}^l_t}{min} \quad \mathcal{L}_t(\mathbf{W}^l_t , \mathbf{e}, \Delta G_t) + \lambda_2 \lVert \mathbf{W}^l_t \rVert_2.
    \label{equ:9}
\end{equation}
Here, we use $L2$ regularization to prevent the model overfitting. The operation of  removing $\mathbf{W}^s_{t-1}$ also corresponds to the best GNN structure search mentioned in Definition~\ref{def:cgl}. Note that only fine-tuning the preserved long-term preference parameters can reduce the computing overload which is of great significance in streaming recommendation setting.

\begin{algorithm}
 \caption{DEGC}
 \label{alg:algorithm}
 \KwIn{A sequence of user-item interaction graphs $\mathcal{G}$.}
 \KwOut{Graph convolution parameters $\mathbf{W}_t$, user embeddings $\mathbf{e}^u_t$, item embeddings $\mathbf{e}^i_t$.}
 \textbf{Process}:\;
\For{each $t$ = 1,2,...,$T$}{
   Initialize the user embeddings with Eqn.~\ref{equ:6} and ~\ref{equ:7}\;
   Train the topmost graph convolution layer $\mathbf{W}^K_t$ with Eqn.~\ref{equ:8}\;
   Obtain $\mathbf{W}^l_{t-1}$ and $\mathbf{W}^s_{t-1}$ with breadth-first search\;
   Refine the $\mathbf{W}^l_{t}$ initialized by $\mathbf{W}^l_{t-1}$ with Eqn.~\ref{equ:9}\;
   Expand the graph convolution layers and train the expanded filters $\Delta \mathbf{W}_t$ with Eqn.~\ref{equ:10}\;
   Prune the expanded filters and finetune the whole model with Eqn.~\ref{equ:11}\;
   Update the $\mathbf{W}_{TA}$ in Eqn.~\ref{equ:5}\;}
\end{algorithm}
\vspace{-0.3cm}
\subsection{Graph Convolution Expanding and Pruning}
\label{sec:gra}
Only the refined long-term preferences are not enough to reflect the users' comprehensive preferences. To extract users' current short-term preferences at segment $t$,  we expand the graph convolution layers and train the expanded part from scratch independently with new data after obtaining fine-tuned $\mathbf{W}^l_{t}$ via operations $1, 2, 3$ in Figure~\ref{fig:overview}. In detail, we add $N$ filters to each graph convolution layer. Then, we initialize the expanded graph convolution parameters $\Delta \mathbf{W}_t$ randomly and train them with $\Delta G_{t}$ while fixing the $\mathbf{W}^l_{t}$:
\begin{equation}
    \underset{\Delta \mathbf{W}_t}{min} \quad \mathcal{L}_t(\mathbf{W}^l_{t};\Delta \mathbf{W}_t , \mathbf{e},\Delta G_t) + \lambda_1 \lVert \Delta \mathbf{W}_t  \rVert_1 + \lambda_g \underset{g}{\sum} \lVert \Delta \mathbf{W}^g_t \rVert_2.
    \label{equ:10}
\end{equation}
Here, we use both $L1$ regularization and group sparse regularization (GSR)~\citep{wen2016learning} to sparsify the expanded convolution parameters $\Delta \mathbf{W}_t$. $g$ is the group consisting of the parameters of each newly added filter. The purpose of sparsification terms here is to prevent the convolution layer width explosion catastrophe if adding $N$ filters to each layer constantly at each segment. 

After obtaining sparsified $\Delta \mathbf{W}_t$ (operation $5$ in Figure~\ref{fig:overview}), we prune $\Delta \mathbf{W}^k_t$ at each layer $k$. Specifically, for each $\Delta \mathbf{W}^k_t$, we first search the filters whose weights are all zeros and remove such filters. Meanwhile, the corresponding parameters in $\Delta \mathbf{W}^{k+1}_t$ of layer $k+1$ are also pruned. In such way, we not only extract the current short-term preferences at segment $t$ but also eliminate the redundant expansion parameters. It needs to be mentioned that our expanding and pruning operations echo the GNN structure optimization in Definition~\ref{def:cgl} again. Finally, the fixed $\mathbf{W}^l_{t}$, the pruned $\Delta \mathbf{W}_t$, and the embeddings $\mathbf{e}$ will be finetuned:
\begin{equation}
    \underset{\mathbf{W}^l_{t}, \Delta \mathbf{W}_t, \mathbf{e}}{min} \quad \mathcal{L}_t(\mathbf{W}^l_{t}; \Delta \mathbf{W}_t, \mathbf{e}, \Delta G_t) + \lambda_1 (\lVert \mathbf{W}^l_{t} \rVert_1 + \lVert \Delta \mathbf{W}_t \rVert_1 ) + \lambda_2 \lVert  \mathbf{e} \rVert_2.
    \label{equ:11}
\end{equation}
The embeddings $\mathbf{e}$ include both user embeddings $\mathbf{e}^u$ and item embeddings $\mathbf{e}^i$. Here, the $L1$ regularization is to sparsify the whole graph convolution structure to facilitate the historical convolution pruning on the next segment $t+1$. The whole workflow of our method is illustrated in Algorithm~\ref{alg:algorithm}. Because our method is actually  orthogonal to previous methods like \textit{experience replay} and \textit{knowledge distillation}, we will combine our method with previous methods to demonstrate its advantages.

%% file: experiments.tex
\section{Experiments}
In this section, we conduct experiments on four real-world time-stamped recommendation datasets to show the effectiveness of our method. We mainly focus on the following questions:

\begin{itemize}[leftmargin=*]

\item \textbf{RQ1:} Whether our method can get better recommendation effects than the state-of-art methods on the most common recommendation metrics like \textit{Recall@20} and \textit{NDCG@20}?

\item \textbf{RQ2:} Whether our method is robust to different datasets and GCN-based recommendation models?

\item \textbf{RQ3:} Whether pruning historical convolution parameters to forget the outdated short-term preferences is necessary for improving streaming recommendation?

\item \textbf{RQ4:} Whether temporal user preference modeling as initialization help learn users' preferences in the GCN update phase?

\end{itemize}
\subsection{Experiment Settings}
\subsubsection{Datasets}
\begin{itemize}[leftmargin=*]
    \item \textbf{Taobao2014\footnote{https://tianchi.aliyun.com/dataset/dataDetail?dataId=46}:} This dataset contains real users-commodities behavior data collected from Alibaba's M-Commerce platforms, spanning 30 days. The rich background informations, like users' location information and behavior timestamp are also included. We filter out users and items with less than 10 interactions.
    \item \textbf{Taobao2015\footnote{https://tianchi.aliyun.com/dataset/dataDetail?dataId=53}:} This dataset contains user behavior data between July 1st, 2015 and Nov. 30th, 2015 accumulated on Taobao.com and the app Alipay. The online actions and timestamps are both recorded. In this work, we only use the first month data for analysis. Besides, We filter out users and items with less than 20 interactions.
    \item \textbf{Neflix\footnote{https://academictorrents.com/details/9b13183dc4d60676b773c9e2cd6de5e5542cee9a}:} This movie rating dataset contains over 100 million ratings from 480 thousand randomly-chosen Netflix customers over 17 thousand movie titles.  The data were collected between October, 1998 and December, 2005 and reflect the distribution of all ratings received during this period. A similar filtering operation is executed and the thresholds are both set as 30. We use the first 24 months data for analysis.
    \item \textbf{Foursquare
    ~\citep{yang2019revisiting, yang2020lbsn2vec++}:} This dataset includes long-term (about 22 months from Apr. 2012 to Jan. 2014) global-scale check-in data collected from Foursquare, a local search-and-discovery mobile APP. The check-in dataset contains 22,809,624 checkins by 114,324 users on 3,820,891 venues. For this dataset, we set the filtering thresholds as 20.
\end{itemize}
The data statistics after filtering of the above datasets are detailed in Appendix~\ref{appendix:dataset}.  Average entity overlapping rate between adjacent segments (AER) is a metric to measure the stability of a data stream. The larger the AER, the more stable the data stream. The data on each segment is split into training, validation, and test sets using a ratio of 8:1:1. We repeat each experiment five times and report the average results to reduce the variance brought by the randomness. 

\subsubsection{Baselines}
\begin{itemize}[leftmargin=*]
    \item \textbf{Finetune}: Finetune first inherits the parameters from the previous segment and then fine-tune the model only with the data of the current segment.  
    \item \textbf{Uniform Sampling (Uniform)}: A kind of naive \textit{experience replay} method which first sample the historical data uniformly and then combine the new data with it. The model is trained with the combined data from scratch at each segment.
    \item \textbf{Inverse Degree Sampling (Inverse) ~\citep{ahrabian2021structure}}: A similar sampling-based \textit{experience replay} method. However, the sampling probability of each interaction is proportional to its user's inverse degree on the interaction graph.
    \item \textbf{ContinualGNN~\citep{wang2020streaming}}: A continual graph learning method which combines the \textit{experience replay} and \textit{knowledge distillation} for existing pattern consolidation.
    \item \textbf{Topology-aware Weight Preserving (TWP)~\citep{liu2020overcoming}}: A \textit{knowledge distillation} method which explores the local structure of the interaction graph and stabilize the parameters playing pivotal roles in the topological aggregation.
    \item \textbf{GraphSAIL~\citep{xu2020graphsail}}: A \textit{knowledge distillation} method which preserves each node's local structure, global structure, and self-information, respectively at the new segment.
    \item \textbf{SGCT~\citep{wang2021graph}}: A \textit{knowledge distillation} method which uses contrastive distillation on each node's embedding where a single user-item graph is used to construct positive samples.
    \item \textbf{MGCT~\citep{wang2021graph}}: A \textit{knowledge distillation} method which uses contrastive distillation on each node's embedding where multiple graphs (user-item, user-user, item-item graphs) is used to construct positive samples.
    \item \textbf{LWC-KD~\citep{wang2021graph}}: Based on MGCT, LWC-KD adds the intermediate layer distillation to inject layer-level supervision.
\end{itemize}
We compare our \textbf{DEGC} model with the above continual graph learning methods. Note that we will show the experiment results of \textbf{DEGC+Finetune} and \textbf{DEGC+LWC-KD} (the combined methods mentioned in Section~\ref{sec:gra}) for a fair comparison. The implementation details are provided in Appendix~\ref{appendix:implementation}. The code is available at \textcolor{blue}{\url{https://github.com/BokwaiHo/DEGC}}.

\subsubsection{Evaluation Metrics}
All the methods are evaluated in terms of \textit{Recall@k} and \textit{NDCG@k}. For each user, our recommendation model will recommend an ordered list of items to her. \textit{Recall@k} (abbreviated as \textit{R@k}) indicates the percentage of her rated items that appear in the top $k$ items of the recommended list. The \textit{NDCG@k} (abbreviated as \textit{N@k}) is the normalized discounted cumulative gain at a ranking position $k$ to measure the ranking quality. Similar to previous related papers~\citep{xu2020graphsail, ahrabian2021structure, wang2021graph}, we set the $k$ as 20.

\begin{table*}[!t]
\caption{The average performance with MGCCF as our base model. $*$ indicates the improvements over baselines are statistically significant ($t$-test, $p$-value $\leq 0.01$).}
\begin{tabular}{|c|cc|cc|cc|cc|}
\hline
\multirow{2}{*}{Method} & \multicolumn{2}{c|}{Taobao2014}            & \multicolumn{2}{c|}{Taobao2015}            & \multicolumn{2}{c|}{Netflix}               & \multicolumn{2}{c|}{Foursquare}                                 \\ \cline{2-9} 
                        & \multicolumn{1}{c|}{Recall@20} & NDCG@20   & \multicolumn{1}{c|}{Recall@20} & NDCG@20   & \multicolumn{1}{c|}{Recall@20} & NDCG@20   & \multicolumn{1}{c|}{Recall@20} & \multicolumn{1}{c|}{NDCG@20}   \\ \hline
Finetune                & \multicolumn{1}{c|}{0.0412}          &   0.0052        & \multicolumn{1}{c|}{0.4256}          &   0.0117        & \multicolumn{1}{c|}{0.3359}          &    0.0580       & \multicolumn{1}{c|}{0.1154}          &         0.0115                       \\ \hline
Uniform                     & \multicolumn{1}{c|}{0.0308}          &      0.0038     & \multicolumn{1}{c|}{0.4194}          &       0.0113    & \multicolumn{1}{c|}{0.3298}          &       0.0533    & \multicolumn{1}{c|}{0.1024}          &       0.0101                         \\ \hline
Inverse                     & \multicolumn{1}{c|}{0.0323}          &     0.0039      & \multicolumn{1}{c|}{0.4217}          &     0.0114      & \multicolumn{1}{c|}{0.3321}          &    0.0536       & \multicolumn{1}{c|}{0.1063}          &        0.0107                        \\ \hline
ContinualGNN            & \multicolumn{1}{c|}{0.0311} & 0.0035 & \multicolumn{1}{c|}{0.4203} & 0.0111 & \multicolumn{1}{c|}{0.3089} & 0.0491 & \multicolumn{1}{c|}{0.1056} & \multicolumn{1}{c|}{0.0098} \\ \hline
TWP                     & \multicolumn{1}{c|}{0.0398}          &    0.0050       & \multicolumn{1}{c|}{0.4320}          &    0.0122       & \multicolumn{1}{c|}{0.3428}          &    0.0580       & \multicolumn{1}{c|}{0.1048}          &  0.0104                              \\ \hline
GraphSAIL               & \multicolumn{1}{c|}{0.0395}          &   0.0051        & \multicolumn{1}{c|}{0.4371}          &   0.0126        & \multicolumn{1}{c|}{\underline{0.3470}}          &  \underline{0.0583}         & \multicolumn{1}{c|}{0.1086}          &      0.0110                         \\ \hline
SGCT                    & \multicolumn{1}{c|}{0.0423}          &    0.0054       & \multicolumn{1}{c|}{0.4411}          & 0.0129          & \multicolumn{1}{c|}{0.3300}          &   0.0572        & \multicolumn{1}{c|}{0.1255}          &      0.0137                          \\ \hline
MGCT                    & \multicolumn{1}{c|}{0.0421}          &   0.0055        & \multicolumn{1}{c|}{0.4446}          &   0.0131        & \multicolumn{1}{c|}{0.3252}          &    0.0562       & \multicolumn{1}{c|}{0.1192}          &       0.0127                         \\ \hline
LWC-KD                  & \multicolumn{1}{c|}{\underline{0.0440}}          &   \underline{0.0059}        & \multicolumn{1}{c|}{\underline{0.4512}}          &   \underline{0.0139}        & \multicolumn{1}{c|}{0.2616}          &   0.0482        & \multicolumn{1}{c|}{\underline{0.1280}}          &              \underline{0.0144}                 \\ \hline
DEGC+LWC-KD             & \multicolumn{1}{c|}{\textbf{\begin{tabular}[c]{@{}c@{}}0.1082*\\ ($\uparrow$ 146\%)\end{tabular}}} & \textbf{\begin{tabular}[c]{@{}c@{}}0.0142*\\ ($\uparrow$ 141\%)\end{tabular}} & \multicolumn{1}{c|}{\textbf{\begin{tabular}[c]{@{}c@{}}0.4892*\\ ($\uparrow$ 8.42\%)\end{tabular}}} & \textbf{\begin{tabular}[c]{@{}c@{}}0.0167*\\ ($\uparrow$ 20.1\%)\end{tabular}*} & \multicolumn{1}{c|}{\textbf{\begin{tabular}[c]{@{}c@{}}0.2949*\\ ($\downarrow$ 15.0\%)\end{tabular}}} & \textbf{\begin{tabular}[c]{@{}c@{}}0.0520*\\ ($\downarrow$ 10.8\%)\end{tabular}} & \multicolumn{1}{c|}{\textbf{\begin{tabular}[c]{@{}c@{}}0.1425*\\ ($\uparrow$ 11.3\%)\end{tabular}}} & \multicolumn{1}{c|}{\textbf{\begin{tabular}[c]{@{}c@{}}0.0178*\\ ($\uparrow$ 23.6\%)\end{tabular}}} \\ \hline
DEGC+Finetune           & \multicolumn{1}{c|}{\textbf{\begin{tabular}[c]{@{}c@{}}0.0825*\\ ($\uparrow$ 87.5\%)\end{tabular}}} & \textbf{\begin{tabular}[c]{@{}c@{}}0.0106*\\ ($\uparrow$ 79.7\%)\end{tabular}} & \multicolumn{1}{c|}{\textbf{\begin{tabular}[c]{@{}c@{}}0.4563*\\ ($\uparrow$ 1.13\%)\end{tabular}}} & \textbf{\begin{tabular}[c]{@{}c@{}}0.0142*\\ ($\uparrow$ 2.16\%)\end{tabular}} & \multicolumn{1}{c|}{\textbf{\begin{tabular}[c]{@{}c@{}}0.3583*\\ ($\uparrow$ 3.26\%)\end{tabular}}} & \textbf{\begin{tabular}[c]{@{}c@{}}0.0612*\\ ($\uparrow$ 4.97\%)\end{tabular}} & \multicolumn{1}{c|}{\textbf{\begin{tabular}[c]{@{}c@{}}0.1324*\\ ($\uparrow$ 3.44\%)\end{tabular}}} & \multicolumn{1}{c|}{\textbf{\begin{tabular}[c]{@{}c@{}}0.0153*\\ ($\uparrow$ 6.25\%)\end{tabular}}} \\ \hline
\end{tabular}
\label{table:mgccf}
\end{table*}

\begin{table}[]
\caption{The average performance with NGCF as our base model. $*$ indicates the improvements over baselines are statistically significant ($t$-test, $p$-value $\leq 0.01$).}
\begin{tabular}{|c|cc|cc|}
\hline
\multirow{2}{*}{Method} & \multicolumn{2}{c|}{Taobao2014}            & \multicolumn{2}{c|}{Netflix}               \\ \cline{2-5} 
                        & \multicolumn{1}{c|}{R@20}      & N@20      & \multicolumn{1}{c|}{R@20}      & N@20      \\ \hline
Finetune                & \multicolumn{1}{c|}{0.0304}          &  0.0040         & \multicolumn{1}{c|}{0.3131}          &     0.0541      \\ \hline
Uniform                     & \multicolumn{1}{c|}{0.0340}          &  0.0038         & \multicolumn{1}{c|}{\underline{0.3263}}          &     0.0525      \\ \hline
Inverse                     & \multicolumn{1}{c|}{0.0347}          &  0.0039         & \multicolumn{1}{c|}{0.3256}          &     0.0518      \\ \hline
ContinualGNN            & \multicolumn{1}{c|}{0.0338} & 0.0036 & \multicolumn{1}{c|}{0.3047} & 0.0479 \\ \hline
TWP                     & \multicolumn{1}{c|}{0.0358}          &   0.0047        & \multicolumn{1}{c|}{0.3159}          &   0.0531        \\ \hline
GraphSAIL               & \multicolumn{1}{c|}{0.0318}          &     0.0042      & \multicolumn{1}{c|}{0.3245}          &   \underline{0.0554}        \\ \hline
SGCT                    & \multicolumn{1}{c|}{0.0350}          &    0.0046       & \multicolumn{1}{c|}{0.3044}          &     0.0533      \\ \hline
MGCT                    & \multicolumn{1}{c|}{0.0346}          &     0.0045      & \multicolumn{1}{c|}{0.2957}          &       0.0511    \\ \hline
LWC-KD                  & \multicolumn{1}{c|}{\underline{0.0380}}          &    \underline{0.0050}       & \multicolumn{1}{c|}{0.2496}          &    0.0454       \\ \hline
DEGC+LWC-KD             & \multicolumn{1}{c|}{\textbf{\begin{tabular}[c]{@{}c@{}}0.0961*\\ ($\uparrow$ 153\%)\end{tabular}}} & \textbf{\begin{tabular}[c]{@{}c@{}}0.0123*\\ ($\uparrow$ 146\%)\end{tabular}} & \multicolumn{1}{c|}{\textbf{\begin{tabular}[c]{@{}c@{}}0.2713*\\ ($\downarrow$ 16.9\%)\end{tabular}}} & \textbf{\begin{tabular}[c]{@{}c@{}}0.0486*\\ ($\downarrow$ 12.3\%)\end{tabular}} \\ \hline
DEGC+Finetune           & \multicolumn{1}{c|}{\textbf{\begin{tabular}[c]{@{}c@{}}0.0816*\\ ($\uparrow$ 115\%)\end{tabular}}} & \textbf{\begin{tabular}[c]{@{}c@{}}0.0107*\\ ($\uparrow$ 114\%)\end{tabular}} & \multicolumn{1}{c|}{\textbf{\begin{tabular}[c]{@{}c@{}}0.3454*\\ ($\uparrow$ 5.85\%)\end{tabular}}} & \textbf{\begin{tabular}[c]{@{}c@{}}0.0594*\\ ($\uparrow$ 7.22\%)\end{tabular}} \\ \hline
\end{tabular}
\label{table:ngcf}
\end{table}

\begin{table}[]
\caption{The average performance with LightGCN as our base model. $*$ indicates the improvements over baselines are statistically significant ($t$-test, $p$-value $\leq 0.01$).}
\begin{tabular}{|c|cc|cc|}
\hline
\multirow{2}{*}{Method} & \multicolumn{2}{c|}{Taobao2014}            & \multicolumn{2}{c|}{Netflix}               \\ \cline{2-5} 
                        & \multicolumn{1}{c|}{R@20}      & N@20      & \multicolumn{1}{c|}{R@20}      & N@20      \\ \hline
Finetune                & \multicolumn{1}{c|}{0.0339}          &     0.0040      & \multicolumn{1}{c|}{0.3179}          &     0.0537      \\ \hline
Uniform                     & \multicolumn{1}{c|}{0.0377}          &   0.0041        & \multicolumn{1}{c|}{\underline{0.3289}}          &    0.0533       \\ \hline
Inverse                     & \multicolumn{1}{c|}{0.0386}          &     0.0042      & \multicolumn{1}{c|}{0.3275}          &     0.0530      \\ \hline
ContinualGNN            & \multicolumn{1}{c|}{0.0382} & 0.0041 & \multicolumn{1}{c|}{0.3035} &  0.0475\\ \hline
TWP                     & \multicolumn{1}{c|}{0.0338}          &     0.0040      & \multicolumn{1}{c|}{0.3204}          &    0.0542       \\ \hline
GraphSAIL               & \multicolumn{1}{c|}{0.0342}          &     0.0042      & \multicolumn{1}{c|}{0.3282}          &    \underline{0.0544}       \\ \hline
SGCT                    & \multicolumn{1}{c|}{0.0342}          &      0.0043     & \multicolumn{1}{c|}{0.3073}          &   0.0519        \\ \hline
MGCT                    & \multicolumn{1}{c|}{0.0357}          &     0.0047      & \multicolumn{1}{c|}{0.2983}          &   0.0516        \\ \hline
LWC-KD                  & \multicolumn{1}{c|}{\underline{0.0402}}          &     \underline{0.0053}      & \multicolumn{1}{c|}{0.2571}          &   0.0461        \\ \hline
DEGC+LWC-KD             & \multicolumn{1}{c|}{\textbf{\begin{tabular}[c]{@{}c@{}}0.0975*\\ ($\uparrow$ 143\%)\end{tabular}}} & \textbf{\begin{tabular}[c]{@{}c@{}}0.0125*\\ ($\uparrow$ 136\%)\end{tabular}} & \multicolumn{1}{c|}{\textbf{\begin{tabular}[c]{@{}c@{}}0.2776*\\ ($\downarrow$ 15.6\%)\end{tabular}}} & \textbf{\begin{tabular}[c]{@{}c@{}}0.0491*\\ ($\downarrow$ 9.74\%)\end{tabular}} \\ \hline
DEGC+Finetune           & \multicolumn{1}{c|}{\textbf{\begin{tabular}[c]{@{}c@{}}0.0833*\\ ($\uparrow$ 107\%)\end{tabular}}} & \textbf{\begin{tabular}[c]{@{}c@{}}0.0109*\\ ($\uparrow$ 106\%)\end{tabular}} & \multicolumn{1}{c|}{\textbf{\begin{tabular}[c]{@{}c@{}}0.3483*\\ ($\uparrow$ 5.90\%)\end{tabular}}} & \textbf{\begin{tabular}[c]{@{}c@{}}0.0596*\\ ($\uparrow$ 9.56\%)\end{tabular}} \\ \hline
\end{tabular}
\label{table:lightgcn}
\end{table}

\subsection{Results and Analysis}
\subsubsection{\textbf{Overall Performance (RQ1)}}
\begin{figure}[ht]
    \centering
    \includegraphics[width=0.48\textwidth]{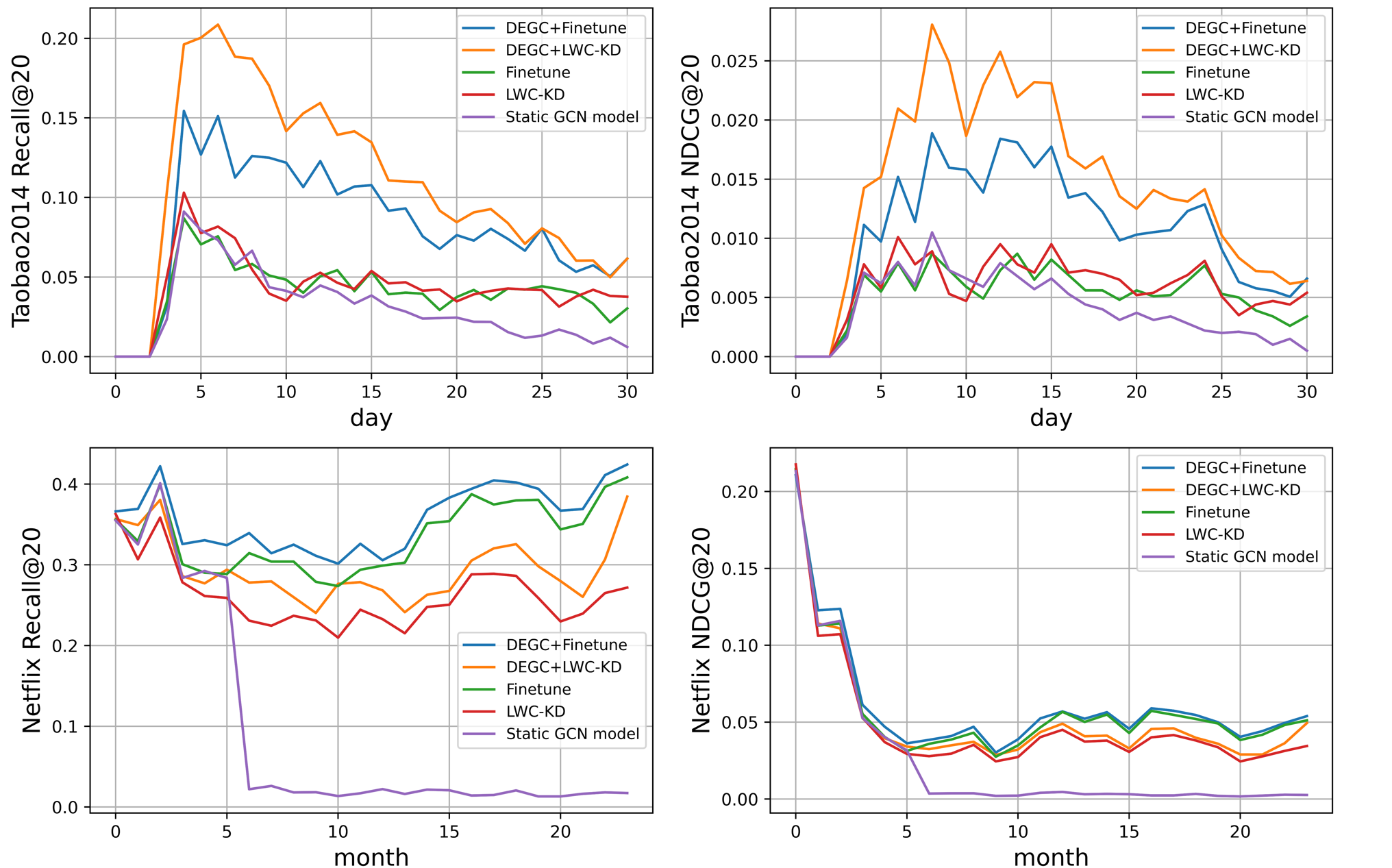}
    \caption{The time-varying model performance on the data stream of Taobao2014 dataset and Netflix dataset.
    }
    \Description{The time-varying model performance on the data stream of Taobao2014 dataset and Netflix dataset.}
    \label{fig:overall_performance}
    \vspace{-0.3cm}
\end{figure}
To answer \textbf{RQ1}, we evaluate our model performance from two perspectives: time-varying performance and average performance on the data stream. In Figure~\ref{fig:overall_performance}, we visualize the \textit{Recall@20} and \textit{NDCG@20} curves on the data streams of Taobao2014 dataset and Netflix dataset. Comparing the DEGC+Finetune and DEGC+LWC-KD with Finetune and LWC-KD, respectively, we can observe that our methods achieve significant time-varying performance gain over their corresponding base methods. Note that the zero performance in the first two days of Taobao2014 dataset is due to the limited amount of data and the model overfitting. The main reason that the observed sharp \textit{NDCG@20} decreases in the first five months of Netflix dataset is the rapidly increasing numbers of users and items. In Table~\ref{table:mgccf}, we show the average performance of different methods on four datasets while choosing MGCCF as the base GCN recommendation model. It can be observed that our DEGC+Finetune and DEGC+LWC-KD get better recommendation effects than all state-of-art methods on all four datasets, except the DEGC+LWC-KD on Netflix. We argue that this is because the poor performance of LWC-KD itself on Netflix. Comparing the DEGC+LWC-KD with LWC-KD independently, our method still improves the performance by $12.7\%$ on \textit{Recall@20} and $7.9\%$ on \textit{NDCG@20}, which also demonstrate the effectiveness of our method. As for the poor performance of SGCT, MGCT, LWC-KD on Netflix, this is resulted by the \textit{over-stability} issue. Such type of \textit{knowledge  distillation} methods can hardly accurately capture the user preferences' shifts  when they change rapidly. Besides, it can be noticed that the performance of \textit{experience replay} methods including Uniform, Inverse, and ContinualGNN are even worse than Finetune. Actually, in the streaming recommendation, such methods can replay previous data containing users' outdated short-term preferences, which negatively influences the model learning on new segments. 

\subsubsection{\textbf{Method Robustness Analysis (RQ2)}}
To answer \textbf{RQ2}, we conduct the experiments with NGCF and LightGCN as the base GCN models on both Taobao2014 and Netflix datasets. The corresponding results are shown in Tables~\ref{table:ngcf} and~\ref{table:lightgcn}. For GCN models NGCF and LightGCN, the improvements of our methods on Taobao2014 are both significant. DEGC+Finetune and DEGC+LWC-KD both achieve the state-of-art recommendation performance. An interesting observation is that DEGC+Finetune in Table~\ref{table:lightgcn} even gets better performance than that in Table~\ref{table:mgccf}. This also shows the performance potential of DEGC on different kinds of base GCN models. As for the dataset Netflix, DEGC+Finetune improves the \textit{Recall@20} by $10.3\%$ and \textit{NDCG@20} by $9.8\%$ over Finetune when choosing NGCF as the GCN model. DEGC+LWC-KD improves the \textit{Recall@20} by $8.7\%$ and \textit{NDCG@20} by $7.0\%$ over LWC-KD, meanwhile. Besides, DEGC+Finetune achieves the best recommendation effect over all previous methods. Similar improvements can also be observed when taking LightGCN as the base GCN model. Such observations demonstrate the robustness of our methods to different datasets and GCN-based recommendation models.

\subsubsection{\textbf{Ablation Study to Historical Convolution Pruning (RQ3)}}
\label{sec:abl hcp}
\begin{figure}[ht]
    \centering
    \includegraphics[width=0.48\textwidth]{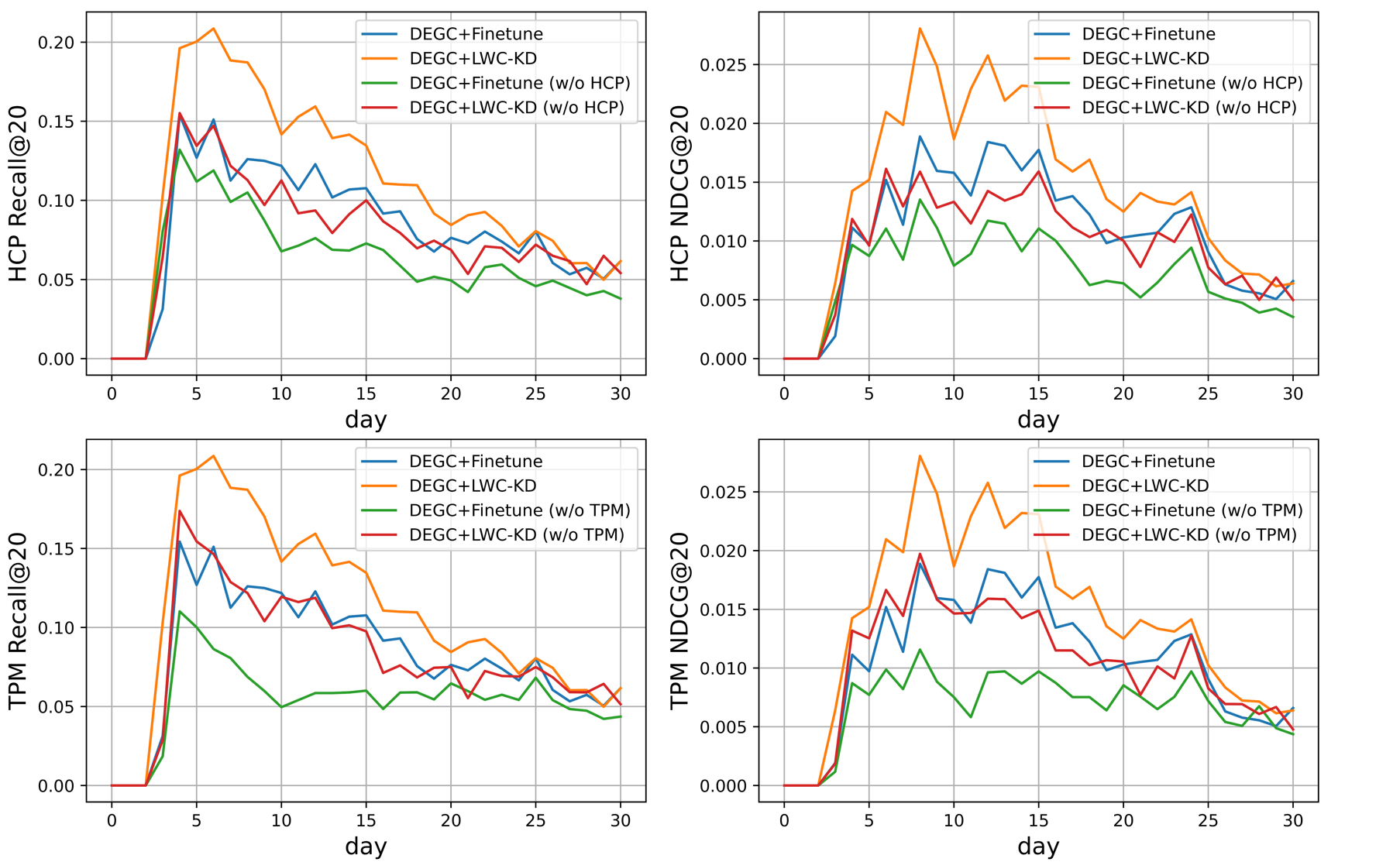}
    \caption{Ablation study to historical convolution pruning (HCP)  and temporal preference modeling (TPM) on Taobao2014 dataset.
    }
    \Description{Ablation study to historical convolution pruning (HCP)  and temporal preference modeling (TPM) on Taobao2014 dataset.}
    \label{fig:ablation}
    \vspace{-0.3cm}
\end{figure}
To answer \textbf{RQ3}, we conduct the experiments with DEGC+Finetune (w/o HCP) and DEGC+LWC-KD (w/o HCP) as the continual graph learning methods while taking the MGCCF as base GCN model on Taobao2014 and Netflix datasets. The top two subfigures of Figure ~\ref{fig:ablation} illustrate the time-varying recommendation performance with two metrics: \textit{Recall@20} and \textit{NDCG@20}. Comparing the DEGC+Finetune with DEGC+Finetune (w/o HCP) and DEGC+LWC-KD with DEGC+LWC-KD (w/o HCP), respectively, we can find that historical convolution pruning is of great significance to our methods' effectiveness. These demonstrate that pruning historical convolution parameter to forget the outdated short-term preferences is necessary. It is also validated that conventional continual graph methods that inherit the parameters learned on the last segment indiscriminately and then finetune them with the new data hinders the model learning on the new segment. This can also be regarded as, at least, part of the reasons that lead to the 'over-stability' challenge in continual learning for streaming recommendation. We also quantitatively analyze the performance drop after removing the historical convolution pruning. We find that DEGC+Finetune decreases by $25.3\%$ and DEGC+LWC-KD decreases by $28.4\%$ on average. The more severe decrease effect of DEGC+LWC-KD is due to that LWC-KD preserves more historical knowledge which may contain users' outdated short-term preferences. We can also observe similar results regarding Netflix dataset in Figure~\ref{fig:ablation_netflix} of Appendix~\ref{appendix:sup ablation}.

\subsubsection{\textbf{Ablation Study to Temporal Preference Modeling (RQ4)}}
To answer \textbf{RQ4}, we conduct the experiments with DEGC+Finetune (w/o TPM) and DEGC+LWC-KD (w/o TPM) on Taobao2014 and Netflix datasets while taking the MGCCF as the base GCN model. From the bottom two subfigures of Figure~\ref{fig:ablation}, we can observe that the \textit{Recall@20} and \textit{NDCG@20} metrics both decrease obviously after removing the temporal preference modeling. This proves that modeling temporal user preference as initialization does benefit the user preference learning in the GCN update phase. Actually, this also corresponds to the other two challenges except the continuous user preference shifts in streaming recommendation: ever-increasing users and intermittent user activities mentioned in Section~\ref{sec:introduction}. Traditional continual graph learning methods like Finetune and LWC-KD directly use the user embeddings learned in the last segment as the embeddings initialization in the current segment. So they can hardly provide an accurate embedding initialization to users whose active intervals on the online platform are longer than a time segment.  Also, they cannot provide a warm embedding initialization for newly coming users. Our temporal preference modeling as user embedding initialization solves such two challenges to some extent and improves the \textit{Recall@20} by $34.3\%$ and $24.9\%$ on average, over DEGC+Finetune (w/o TPM) and DEGC+LWC-KD (w/o TPM), respectively. The similar trends on Netflix dataset can also be observed in Figure~\ref{fig:ablation_netflix} of Appendix~\ref{appendix:sup ablation}.



%% file: conclusion.tex
\section{Conclusion}
Streaming recommendation has attracted great attention due to the dynamics of the real world. 
In this paper, we first propose the temporal preference modeling as the user embedding initialization of each time segment. Then, we start from the \textit{model isolation} perspective and propose the \textit{historical graph convolution pruning and refining} and \textit{graph convolution expanding and pruning} operations, in such ways to only preserve useful long-term preferences and further extract current short-term preferences. Extensive experiments on four real-world datasets and three most representative GCN-based recommendation models also demonstrate the effectiveness and robustness of our method. 

%% file: appendix.tex
\section{Notations}
\label{appendix:notations}
We summarize the main notations used in this paper in Table~\ref{table:notation}.

\begin{table}[h]
\caption{Major notations.}
\begin{tabular}{c l}
\hline
$T$     & \begin{tabular}[c]{@{}l@{}}The total number of data/time segments\end{tabular} \\ 
$K$     & \begin{tabular}[c]{@{}l@{}}The total number of graph convolution layers\end{tabular} \\ 
$\widetilde{\textbf{D}}$    & \begin{tabular}[c]{@{}l@{}}The user-item interaction data stream\end{tabular} \\ 
$D_t$     & \begin{tabular}[c]{@{}l@{}}The data streaming into the system at time segment $t$\end{tabular} \\ 
$\Delta G_t$     & \begin{tabular}[c]{@{}l@{}}The graph structure of interaction data $D_t$\end{tabular} \\ 
$G_t$     & \begin{tabular}[c]{@{}l@{}} The graph structure of the union of $D_1,D_2,...,D_t$\end{tabular} \\ 
$\mathbf{S}_{t}$     & \begin{tabular}[c]{@{}l@{}}The graph convolution structure at segment $t$\end{tabular} \\ 
$\mathbf{W}_{t}$     & \begin{tabular}[c]{@{}l@{}}The graph convolution parameters at segment $t$\end{tabular} \\ 
$\mathbf{W}^K_{t}$     & \begin{tabular}[c]{@{}l@{}}The topmost graph convolution layer parameters at segment $t$\end{tabular} \\ 
$\mathbf{W}^s_{t}$     & \begin{tabular}[c]{@{}l@{}}The short-term preference-related parameters of $\mathbf{W}_{t}$ \end{tabular} \\ 
$\mathbf{W}^l_{t}$ & \begin{tabular}[c]{@{}l@{}} The long-term preference-related parameters of $\mathbf{W}_{t}$\end{tabular} \\ 
$\Delta \mathbf{W}_{t}$     & The expansion part of graph convolution at segment $t$\\ \hline
\end{tabular}
\label{table:notation}
\end{table}

\section{Data statistics of filtered datasets}
The data statistics of fours filter datasets used in this work are summarized in Table~\ref{table:dataset}.
\label{appendix:dataset}
\begin{table}[]
\caption{Data statistics of filtered datasets.}
\begin{tabular}{|c|c|c|c|c|}
\hline
Dataset                                                                                                  & Tb2014 & Tb2015 & Netflix & Foursquare \\ \hline
user \#                                                                                                   &    8K        &     192K       &    301K     &    52K         \\ \hline
item \#                                                                                                   &      39K     &     10K        &     9K    &   37K         \\ \hline
interaction \#                                                                                          &      749K      &      9M      &   49M      &   2M         \\ \hline
time span                                                                                                    &     31 days       &     123 days      &  74 months       &     22 months      \\ \hline
\begin{tabular}[c]{@{}c@{}}AER \end{tabular}     &  35.5\%          &   26.0\%         &     58.4\%    &       60.0\%      \\ \hline
\end{tabular}
\label{table:dataset}
\end{table}

\section{Supplementary Ablation Study}
\begin{figure}[ht]
    \centering
    \includegraphics[width=0.50\textwidth]{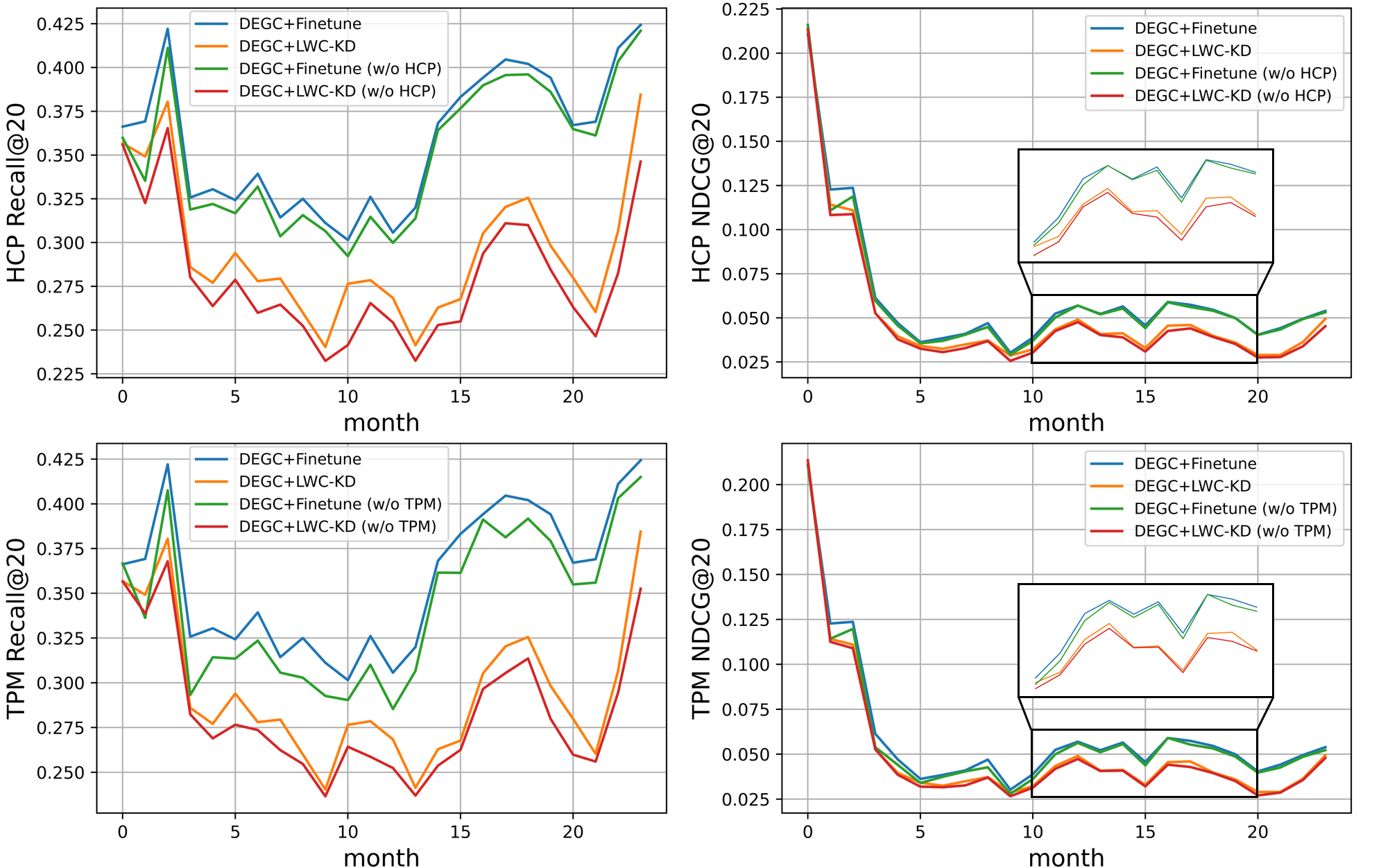}
    \caption{Ablation study to historical convolution pruning (HCP) and temporal preference modeling (TPM) on Netflix dataset.
    }
    \Description{Ablation study to historical convolution pruning (HCP) and temporal preference modeling (TPM) on Netflix dataset.}
    \label{fig:ablation_netflix}
    \vspace{-0.3cm}
\end{figure}

The ablation study results on Netflix dataset are present in Figure~\ref{fig:ablation_netflix}.
\label{appendix:sup ablation}

\section{Implementation Details}
\label{appendix:implementation}
 We report our implementation details here. We use the Adam optimizer with an initial learning rate 
as 0.001. The embedding size and the width of each graph convolution layer are set to 128. The $L1$  regularization coefficient $\lambda_1$ is set to 0.001. The $L2$ regularization coefficient $\lambda_2$ and the GSR regularization coefficient $\lambda_g$ are set to 0.01. We set the batch size as 1,000 when training the GCN models. The number $N$ of the expansion filters at each layer is set as 30. Without specifications, the hyper-parameters are set same as the original papers. We implement our algorithm with Tensorflow and test it on the NVIDIA GeForce RTX 3090 GPU with 24 GB memory.